\documentclass[ reprint,onecolumn,11pt, nofootinbib]{revtex4-1}
\usepackage{graphicx}\usepackage{color}
\usepackage{amsmath}
\usepackage{stmaryrd}

\def\bea{\begin{eqnarray}}
\def\eea{\end{eqnarray}}
\begin{document}

\title{Light dark matter scattering in outer neutron star crusts}

\author{Marina Cerme\~no $^1$~\footnote{marinacgavilan@usal.es}, M. \'Angeles P\'erez-Garc\'ia$^1$~\footnote{mperezga@usal.es} and Joseph Silk$^{2, 3, 4}$~\footnote{silk@iap.fr}}

\affiliation{$^1$ Department of Fundamental Physics, University of Salamanca, Plaza de la Merced s/n 37008 Spain\\ $^2$Institut d'Astrophysique,  UMR 7095 CNRS, Universit\'e Pierre et Marie Curie, 98bis Blvd Arago, 75014 Paris, France\\ $^3$Department of Physics and Astronomy, The Johns Hopkins University, Homewood Campus, Baltimore MD 21218, USA\\
$^4$Beecroft Institute of Particle Astrophysics and Cosmology, Department of Physics, University of Oxford, Oxford OX1 3RH, UK,\\
}

\date{\today}

\begin{abstract}

We calculate for the first time the phonon excitation rate in the outer crust of a neutron star due to  scattering from light dark matter (LDM) particles gravitationally boosted into the star. We consider  dark matter particles in the sub-GeV mass range scattering off a periodic array of nuclei through an effective scalar-vector interaction with nucleons. We find that LDM effects cause a modification of the net number of phonons in the lattice as compared to the standard thermal result. In addition, we estimate the contribution of LDM to the ion-ion thermal conductivity in the outer crust and find that it can be significantly enhanced at large densities. Our results imply that for magnetized neutron stars the LDM-enhanced global 
conductivity in the outer crust will tend to  reduce the anisotropic heat conduction between perpendicular and parallel directions to the magnetic field.

\end{abstract}

\maketitle

\section{Introduction}

Dark matter constitutes the most abundant type of matter in our universe and its density is now experimentally well-determined $\Omega_{CDM}h^2= 0.1199 \pm 0.0027$ \cite{cdm}. Worldwide efforts to constrain its nature and interactions have led the community  to a puzzling situation where null results coexist with direct detection experiments that find high significance excesses \cite{dama}. In particular, in the low mass region of DM candidates i.e. $m_{\chi}<1\; \rm GeV/c^2$, cosmological, astrophysical and collider constraints  seem to be the most important, see for example a discussion in \cite{lin}. Direct detection searches of thermalized galactic DM are mostly based on nuclear recoils on selected targets. In this scenario, LDM particles with masses much smaller than that of the nucleon, $m_\chi \ll m_N$, can only provide energies $\sim$ eV which are below the $\sim$keV threshold for conventional terrestrial searches. If one, instead, considers LDM scattering off bound electrons, energy transfer can cause excitation or even ionization and thus seems promising for exploring  the phase space in a complementary way in the near future \cite{ele}.
DM hitting terrestrial targets is expected to have low velocities $v_\chi \sim 10^{-3}$ (we use $c=\hbar=1$ units) as the gravitational boost is small for the Earth i.e. its Lorentz factor $\gamma_{\varoplus} =1/\sqrt{1-\beta^2_\varoplus} \sim 1$  with $\beta_\varoplus=\sqrt{\frac{2GM_\varoplus}{R_\varoplus}}$. However, for compact objects, such as neutron stars (NSs) with masses $M_{NS}\simeq 1.5 M_\odot$ and radius  $R_{NS}\simeq 12$ km, $\gamma_{NS}\sim 1.26$ or $v_\chi \sim 0.7$ and thus provides a mechanism to boost particles to higher velocities or, accordingly, test the same length scales with smaller projectile masses. In particular, the outer crusts 
in NSs are formed by periodically arranged nuclei with typical densities ranging from $\rho \simeq  2 \,10^6-4\,10^{11}$ $\rm g/cm^3$. In the single-nucleus description \cite{chamel}, a series of nuclei with increasing baryonic number, $A$, from Fe to Kr form a lattice before neutrons start to leak out of nuclei. At these high densities, electrons form a degenerate Fermi sea. At even larger densities and up to nuclear saturation density, around $\rho_0\simeq n_0 m_N \simeq 2.4\,10^{14}$ $\rm g/cm^3$, a number of different nuclear structures called {\it pasta} phases appear \cite{pasta}.

In this work we study the effect of LDM scattering in the production of quantized lattice vibrations (phonons) in the outer NS crust. Later, we will discuss  how this result can impact subsequent quantities of interest, such as the ion thermal conductivity, that are relevant for computing the cooling behavior of NSs. Phonons are quantized vibrational modes characterized by a momentum $\vec{k}$ and polarization vector $\vec{\epsilon}_{\lambda}$ appearing in a nuclear periodic system \cite{ziman}. They can have a number of different sources.  They can be excited due to non-zero temperature $T$  in the medium. The Debye temperature allows us to evaluate the importance of the ion motion quantization. For a bcc lattice \cite{carr}, for example, $T_D\simeq 0.45 T_p$, being $T_p=\omega_p /k_B=\sqrt{\frac{4 \pi n_AZ^2e^2}{k_B^2 m_A}}$ the plasma temperature associated to a medium of ions with number density $n_A$, baryonic number $A$, electric charge $Ze$ and mass $m_A$. $k_B$ is the Boltzmann constant.
At low temperatures $T<T_D$, the quantization becomes increasingly important and the thermal phonons produced are typically acoustic modes, following a linear dispersion relation $\omega_{{k},\lambda}=c_{l, \lambda}|{\vec k}|$, where $c_l=\frac{\omega_p/3}{ (6\pi^2 n_A)^{1/3}}$ is the sound speed. 
In addition, phonon production can be caused by an external scattering agent, for example, standard model neutrinos.  In this respect, weak probes such as cosmological neutrinos with densities $n_\nu \sim 116$ $\rm cm^{-3}$ per flavor have been shown to provide small phonon production rates in a crystal target \cite{ferreras}. Due to the tiny mass of the neutrino, the experimental signature of this effect seems however hard to confirm. 
The main interest in the astrophysical context we discuss in this contribution follows as phonon excitation in a periodic system, such as the outer crust of a NS, can affect the thermal transport coefficients in the star. The potential modification of transport properties of heat/energy in the external layers in NSs is crucial to possibly identifying relevant distortions in the cooling behavior of these astrophysical objects in rich LDM environments.

The structure of this contribution is as follows. In section II, we present the effective field theory Lagrangian model using dark matter-nucleon contact interactions via  scalar and vector couplings in a relativistic framework and compute the single phonon excitation rate, discussing sources of uncertainty. Later, in section III we compute the thermal conductivity in the outer crust with LDM contributions comparing the results to the standard thermal value and discussing possible astrophysical consequences. Finally, in Section IV we give our conclusions.

\section{Light Dark Matter scattering and phonon excitation rate}

 In this work, we consider the interaction of an incoming fermionic DM particle, $\chi$, scattering quasi-elastically with a nucleus in the outer NS crust lattice via scalar and vector couplings \cite{haxton, cermeno} composed of  $Z$ protons $(p)$ and $A-Z$ neutrons $(n)$
\begin{equation}
\mathcal{L_I}=\sum_{N=n,p}g_{s,N}\chi\overline{\chi} N \overline{N}+g_{v,N}\chi\gamma^{\mu}\overline{\chi} N\gamma_{\mu} \overline{N},
\label{int_l}
\end{equation}
where $g_{s,N}$ $(g_{v,N})$ are the effective scalar (vector) couplings of the DM particle to the nucleon ($N$) field. We will focus on a weakly interacting candidate (WIMP) with mass  in the sub-GeV range. This interaction is equivalent to considering a Fermi four-fermion interaction model, where the effective couplings of mass dimension $(-2)$ for these operators are obtained by integrating out the propagator of a generic $\phi$ mediator with mass $M_\phi$.  Let us mention that, indeed, more effective operators for Dirac LDM candidates are possible, see for example, Table I in \cite{tait} for leading  coupling contributions to Standard Model fermions. However, in order to keep our description concise for the sake of clarity, we will restrict here to the spin-independent interaction model used in previous works \cite{cermeno}. 

Motivated by the need to compare bounds from colliders to direct detection, we describe interactions of DM with quarks $q=u,d$ and averaging in terms of nucleon fields we can write for the vector case $g_{v,N}/M_\phi^2\sim 1/{\Lambda_v}^2$ and $g_{s,N}/M_\phi^2\sim m_q/{\Lambda_s}^3$ where ${\Lambda_v}$ ( ${\Lambda_s}$) is the suppression mass scale for the vector (scalar) case, assuming the effective couplings are of order $O(1)$ and can be absorbed into ${\Lambda_{s,v}}$ \cite{ci}. Using bounds from CMS and ATLAS \cite{limits} we set  ${\Lambda_v}\gtrsim 1$ TeV and ${\Lambda_s}\gtrsim 100$ GeV.

 We denote $p'^{\mu}_N=(E'_N, \vec{p'_N})$ and $p^{\mu}_N=(E_N, \vec{p_N})$ as the four momentum for the outgoing and incoming nucleon, and $p'^{\mu}_\chi=(E'_\chi, \vec{p'_\chi})$ and $p^{\mu}_\chi=(E_\chi, \vec{p_\chi})$ those analogous for the LDM particle, respectively. Momentum transfer is denoted by $q^{\mu}=p'^{\mu}_\chi-p^{\mu}_\chi$. 

Generically, given an interaction potential $\mathcal {V}$ felt by an interacting DM particle when approaching a nucleus in the periodic lattice, the single phonon excitation rate {\it per mode} can be obtained using the Fermi golden rule, $R_{\vec{k},\lambda}={2\pi}\delta (E_f-E_i)|\langle f |\mathcal {V}|i\rangle |^2$ where $i$ and $f$ are the initial and final states considered and $\delta (E_f-E_i)$ assures energy conservation. Given the fact that incoming (outgoing) LDM particles suffer a very moderate perturbation from the plane wave state, we will describe its incoming (outgoing) quantum state as  $\Psi_{\vec{p_{\chi}}}(\vec{r})=\frac{1}{\sqrt{V}}e^{i\vec{p_{\chi}}\vec{r}}$ with $V$ the volume of the system.
The interaction potential felt by the LDM particle is the sum \cite{ferreras} over lattice sites
%
%
$\mathcal{V}(\vec{r})=\sum_{j}v(\vec{r}-\vec{r_j})$ that we describe for the sake of simplicity as impenetrable point-like spheres $v(\vec{r}-\vec{r_j})=\delta^3(\vec{r}-\vec{r_j})\; v_0$. We, nevertheless, comment on corrections to this picture later in the manuscript. 

Using the Born approximation, the scattering amplitude for an incident $\chi$ particle can be written as
\begin{equation}
f(\vec{p_\chi},\vec{p'_\chi})\simeq -\frac{m_{\chi}}{2\pi}\int e^{i(\vec{p_\chi}-\vec{p'_\chi})\vec{r'}}v(\vec{r'})d^3\vec{r'},
\label{born1}
\end{equation} 
%
and from its squared value, the differential cross section in the center of mass frame,  $\frac{d\sigma}{d\Omega}|_{CM}=|f(\vec{p_{\chi}},\vec{p'_{\chi}})|^2$. The validity of the Born approximation is  provided by the finite-range  potential $\mathcal{V}(\vec{r})$ so the condition $|(\vec{p_{\chi}}-\vec{p'_{\chi}}).\,\vec{r'}|\ll1$ is fulfilled, being $|\vec{r'}|$ a typical target size. The effective interaction potential can be obtained from the squared interaction matrix element as calculated from the  Lagrangian in eq.(\ref{int_l}) as $\frac{d\sigma}{d\Omega}|_{CM}=\frac{|\mathcal{\overline{M}}_{\chi N}|^2}{64\pi^2 s}$. First, we compute the scattering amplitude $|\mathcal{\overline{M}}_{\chi N}|^2$ being $s=(p_N+p_\chi)^2$ the Mandelstam variable. Adding the contribution over proton and neutron amplitudes coherently, we can obtain the LDM particle-nucleus differential cross-section and then integrate to find a relation between the total  cross-section $\sigma_{\chi A} \simeq 4\pi a^2$ or, equivalently, the effective potential from eq.(\ref{born1}), and the scattering length, $a$, at low incident energies. We  obtain $v(\vec{r})=\frac{2\pi a}{m_{\chi}}\delta (\vec{r})$. Besides, we have used a normalization of the delta function as $\int_{VT} \delta(x) d^4x=1$. From the  Lagrangian in eq.(\ref{int_l}) the spin-averaged scattering amplitude \cite{cermeno} reads
%
\begin{equation}
\begin{aligned}
|\mathcal{\overline{M}}_{\chi \rm N}|^2 =& 4g_{s,N}^2 [({p_N}{p'_N}+m_N^2)({p_\chi}{p'_\chi}+m_{\chi}^2)] + 8 g_{v,N}^2 [2m_N^2m_{\chi}^2 -\\
& m_N^2p'_\chi p_\chi-m_{\chi}^2 p_N p'_\chi+(p'_\chi p'_N)(p_N p_\chi)+(p'_Np_\chi)(p_Np'_\chi)]+ \\
& 8 g_{s,N} g_{v,N} [m_N m_{\chi}(p_N+p'_N)(p_\chi+p'_\chi)].
\end{aligned}
\end{equation}
\normalsize
Due to the mildly relativistic nature of nucleons inside the nucleus, energy and momentum will lie close to the Fermi surface values $E_{FN}, |\vec{p}_{FN}|$ and $ |\vec{p}_{N}|^{2} \sim |\vec{p'}_{N}|^{2} \sim  |\vec{p}_{FN}|^{2} \ll m^2_N$. We will approximate the product $p'_N p_N=E_N E'_N-|\vec{p_N}||\vec{p_N}'|cos\; \theta_{\vec{p_N},\vec{p_N}'}\simeq E_{FN}^2$. On the other hand, for the more relativistic DM particle products $p_\chi p'_\chi=E_\chi E'_\chi-|\vec{p_\chi}||\vec{p'_\chi}| cos \; \theta_{\vec{p_\chi},\vec{p'_\chi}}=E_{\chi}^2-|\vec{p_{\chi}}|^2cos \; \theta_{\chi}=m_{\chi}^2+|p_{\chi}|^2(1-cos \; \theta_{\chi})$ where we use $\theta_{\vec{p_\chi},\vec{p_\chi}'}\equiv \theta_{\chi}$. 
%
The density dependence will be retained using a parametrization of the nuclear Fermi momentum $|\vec{p}_{FN}| \sim(3 \pi^2 n_0 Y_N)^{1/3}$ and the nuclear fractions $Y_p=Z/A$, $Y_n=(A-Z)/A$. If we now average over angular variables,
\begin{equation}
\begin{aligned}
\int_{-1}^{1} 2\pi d( cos \; \theta_{\chi}) |\mathcal{\overline{M}_{\chi \rm N}}|^2 =&  16\pi g_{s,N}^2 [(2m_N^2+|\vec{p}_{FN}|^2)(2m_\chi^2+|\vec{p_{\chi}}|^2)]+ \\
& 32\pi g_{v,N}^2 [2E_{\chi}^2E_{FN}^2-m^2_N|\vec{p_{\chi}}|^2-m_{\chi}^2|\vec{p}_{FN}|^2] +\\ 
& 128 \pi g_{s,N}g_{v,N}[m_Nm_{\chi}E_{FN}E_{\chi}].
\end{aligned}
\label{eq1}
\end{equation}
\normalsize
In the nucleus, we can use the previous expression, eq. (\ref{eq1}), to find the coherent contribution of the $A$ nucleons in a similar way to what is done in direct detection \cite{ddetec},

%
\begin{equation}
\int_{-1}^{1} 2\pi d ( cos \; \theta_{\chi}) |\mathcal{\overline{M}_{\chi A}}|^2 \simeq {m^2_A} \left (\frac{Z}{m_p}\sqrt{|\tilde{\mathcal{M}_p}|^2}+\frac{(A-Z)}{m_n}\sqrt{|\tilde{\mathcal{M}_n}|^2}\right) ^2,
\end{equation}
%
with  $\int_{-1}^{1} 2\pi d( cos \; \theta_{\chi}) |\mathcal{\overline{M}_{\chi \rm N}}|^2 \equiv  |\tilde{\mathcal{{M}_{\rm N}}}|$. The Mandelstam variable $s=(p_A+p_\chi)^2=m_{\chi}^2+m_A^2+2E_AE_{\chi}-2\vec{p_A}\vec{p_{\chi}}$ can be approximated as $s\simeq (m_\chi + m_A)^2$, neglecting the mildly relativistic nuclei momenta. Thus we can express the cross-section in the center of mass frame as
\begin{equation}
\sigma_{A,\chi}= 4 \pi a^2= {m^2_A} \frac{\left(\frac{ Z}{m_p}\sqrt{|\tilde{\mathcal{M}_p}|^2}+\frac{(A-Z)}{m_n}\sqrt{|\tilde{\mathcal{M}_n}|^2}\right) ^2}{16 \pi (m_\chi+ m_A)^2}.
\label{se}
\end{equation}
From a zero-order momentum expansion, we recover the usual expression for direct detection spin independent cross-section  at low energies \cite{gluscev}  for each coupling $\sigma_{A,\chi}\rightarrow \frac {\mu^2_{\chi A}}{\pi} (Z  g_{s,p}+ (A-Z)g_{s,n})^2$ where $\mu_{\chi A}=\frac{m_\chi m_A}{m_\chi+ m_A}$ is the reduced $\chi-A$ mass.  Note at this point that the $\sim A^2$ enhancement in the obtained cross-section remains as the coherence condition $\lambda \ge R_{A}$ is fulfilled, being $R_A$ the nuclear radius and $\lambda=h/|\vec{p_\chi}|$ the De Broglie wavelength. In addition, the contribution of the nuclear lattice will be described by the summations extended over the lattice sites or, equivalently, by the inclusion of the structure factor $S(q)\sim |\sum_{j} e^{-i\vec{q}\vec{r_j}}|^2$, in the full phonon excitation rate expression  as will be shown later in the manuscript. Some studies have included form factors $F^2(q)$ to correct a point-like nucleus nature approach \cite{gluscev}, however since we will be focusing on $q\rightarrow 0$ limit we will consider them as unity for the sake of simplicity. In what follows we will refer to $\vec{p'}\equiv \vec{p'_\chi}$ and $\vec{p} \equiv \vec{p_\chi}$. The single-phonon excitation time rate from the ground state now reads 
\begin{equation}
R^{(0)}_{{k},\lambda}=\frac{4\pi^2a^2}{V^2m_{\chi}^2} \delta (E_{\vec{p'}}+\omega_{{k},\lambda}-E_{\vec{p}})\; 2\pi \; |\sum_{j} \langle 1, \vec{k} \lambda |e^{-i\vec{q}\vec{r_j}}|0 \rangle|^2,
\label{rate2}
\end{equation}
where $\vec{r}_j=\vec{x}^{(0)}_j+\vec{u}_j$ with $\vec{x}^{(0)}_j$ the lattice point and $\vec{u}_j$ the displacement vector \cite{Ashcroft}. We must note at this point that the previous expression includes the squared modulus of the Fourier transform of the periodically arranged lattice sites including thus the usual description in terms of the structure factors $S(q)$. This function provides information on the spatial distribution through a  correlation function and presents maxima at the crystal nuclear positions. The contribution of this factor to the global cross-section describes  coherent scattering from all of the different nuclei as discussed in \cite{horo}. There, the effect of efficient low-energy scattered WIMPS from the interior of the stelar DM distribution was mentioned as an additional factor to prevent  DM escaping from the NS once inside. In this way it thus constitutes a mechanism for {\it trapping} DM, besides the deep gravitational potential felt by these sub-GeV mass particles.


Beyond this point, we will consider an isotropic medium and since the Born approximation  $|\vec{q}.{\vec r'}|\ll1$ holds, it is most likely that acoustic modes are excited. It follows that
\begin{equation}
-i\vec{q}\sum_{j} e^{-i\vec{q}\vec{x}^{(0)}_j} \langle 1, \vec{k} \lambda |\vec{u}_j|0 \rangle = -i n_A  \delta^{(3)}(\vec{k}-\vec{q})  \sqrt{\frac{|\vec{k}|}{2m_Ac_l }},
\end{equation}
where we have used the continuum limit $\sum_j \rightarrow n_A \int d^3x$ and the fact that  all $\vec{k}$ have a polarization vector that verifies $\vec{\epsilon_l} // \vec{k}$ and the other two vectors are perpendicular to $\vec{k}$. 
Finally eq.\eqref{rate2} can be written as
\begin{equation}
R^{(0)}_{{k}}=\frac{4\pi^2a^2}{m_{\chi}^2V} \delta (E_{\vec{p'}}+\omega_{{k},\lambda}-E_{\vec{p}})\; 2\pi n^2_A\frac{ \delta^{(3)}(\vec{k}-\vec{q}) |\vec{k}|}{2m_Ac_l}.
\end{equation}

At this point we must consider the peculiarities of the incoming LDM phase space distribution $f_{\chi}(\vec{p})$ as it will impact the averaged final phonon excitation rate.  Typically, for the Sun or the Earth the uncertainties have different sources including orbital speed of the Sun, escape velocity from the DM halo and the form of the phase space distribution itself. About the latter and in local searches, direct and indirect detection  are affected in different manner. For example, direct detection is sensitive to  DM with high velocities \cite{dir} while for indirect detection the low-velocity part of the distribution is tested \cite{ind, choi} . 

A popular choice is obtained using an approximation based on an isotropic sphere with density profile $\rho_{DM}(r) \propto r^{-2}$ of collisionless particles, i.e. a Maxwell-Boltzmann type with a local mass density $\rho_{LDM} = 0.3$ $\rm GeV$ $\rm cm^{-3}$. Uncertainties on the knowledge of the distribution function must be carefully considered as this impacts accuracy when translating event rates to constraints on particle physics models of DM.

In the case we analyse here of a more compact object, it is the high velocity part of the distribution that is tested, as typical values for boosted root-mean-squared velocities are ${<v^2>}\sim {2 G M_{NS}/R_{NS}}\sim (0.6)^2$. 
For these relativistic regimes one must use the Maxwell-J\"uttner distribution \cite{juttner} function and, more properly, take into account the space time curvature due to gravitational field created by the NS \cite{Kremer}
\begin{equation}
f_{\chi}(\vec{p})=\frac{n_\chi \mu }{4 \pi m_\chi^3 K_2(\mu)} e^{-\mu \sqrt{1+g_1(r)\frac{|\vec{p}|^2}{m_\chi^2}}},
\label{f}
\end{equation}
where $\mu=\frac{m_\chi}{k_BT}$ and  $K_2(\mu)$ is the modified Bessel function of second kind defined as  $K_n(\mu)=\left( \frac{\mu}{2}\right)^n \frac{\Gamma(1/2)}{\Gamma(n+1/2)} \int_0^\infty e^{-\mu y}(y^2-1)^{n-1/2}dy$. The isotropic Schwarzschild metric for the gravitational field created by the NS source is \cite{Kremer} $ds^2=g_0(r)(dx^0)^2-g_1(r) \delta_{ij} x^i x^j$, $i,j=1,2,3$. 

Note that in the close vicinity of the NS where we will be interested in assesing our quatities of interest, $r\sim R_{NS}$, and follows that  $g_1(R)=\left( 1+ \frac{GM_{NS}}{2R_{NS}}\right)^4\sim 1.42$, $g_0(R)=\left( 1- \frac{GM_{NS}}{2R_{NS}}\right)^2/\left( 1+ \frac{GM_{NS}}{2R_{NS}}\right)^2\sim 0.69$. The distortion from the flat space with a Minkowski metric efectively sets $g_0(r), g_1(r) \neq 1$ as expected. Furthermore, if we obtain from the above the root-mean-squared $\sqrt{<v^2>}\sim 0.6$ this implies $\mu \approx 6.7$ \cite{Hakim, Cercignani}.   

The normalization condition is such that the particle four-flow $J^\alpha$ can be defined and taking the $\alpha=0$  component we obtain $\int d^3\vec{p}f_{\chi}(\vec{p})\sqrt{-g}/g_0=J^0=n_{\chi}/\sqrt{g_0}$ with  $\sqrt{-g}=\sqrt{g_0 g^3_1}$. $n_{\chi}$ is the DM number density near the NS.  Note that at non-relativistic velocities and flat space we do recover the  Maxwell-Boltzmann distribution as expected.   Further, we consider all outgoing $\chi$ states are allowed as the net number will be tiny as compared to ordinary matter. The  phonon excitation time rate must be weighted with the momenta of the {\it local} $\chi$ phase space that, as mentioned, is shifted to the relativistic values
%
\begin{eqnarray}
R^{(0)}_{{k}} \nonumber & = & \frac{4\pi^3n_A^2V }{m_{\chi}^2m_Ac_l} \int \frac{d^3\vec{p} \,f_{\chi}(\vec{p})}{(2 \pi)^3}  \int \frac{d^3\vec{p'}}{(2 \pi)^3}\delta^{(3)}(\vec{k}-\vec{q})\; \delta (E_{\vec{p'}}+\omega_{\vec{k},\lambda}-E_{\vec{p}})|\vec{k}|a^2.\\ 
\label{rate31}
\end{eqnarray}
\normalsize
Computing the zeros of the delta function and expressing the incoming momentum as $|\vec{p}_{0}|= \sqrt{\gamma^2-1}m_{\chi}$  we obtain an interval of kinematically allowed $|\vec{k}|$ values $0\leq |\vec{k}|\leq 2m_{\chi} \left(  \frac{c_l\gamma }{(c_l^2-1)}+\frac{\sqrt{\gamma^2-1}}{|c_l^2-1|}\right)$ and the eq.\eqref{rate31} takes the form,

%

\begin{equation}
R^{(0)}_{{k}}=\frac{8\pi^4n^2_A V}{(2 \pi)^6 m_{\chi}^2m_Ac_l} \int_{0}^{\infty} |\vec{p}|^2d|\vec{p}| f_{\chi}(\vec{p})\frac{|\gamma m_{\chi}-|\vec{k}|c_l|}{m_{\chi}\sqrt{\gamma^2-1}}a^2.
\end{equation}

\begin{figure}[ht]
\centering
\includegraphics[width=0.5\textwidth, angle=0,scale=1.25]{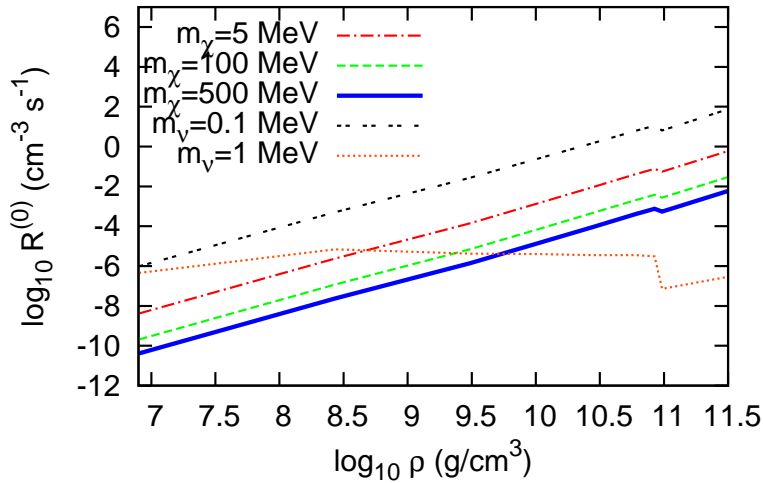}
\caption{Averaged single phonon excitation rate per unit volume as a function of density in the outer crust. DM particle masses  $m_\chi=500, 100$ and $5$ MeV are used and  $n_\chi/n_{0,\chi}=10$. Neutrino contribution at $|\vec{k}|\rightarrow 0$, $R_{\nu 0}$, is also shown for $m_{\nu}=0.1, 1$ eV. See text for details. }
\label{fig1}
\end{figure}

In Fig.(\ref{fig1}) we show the single phonon excitation rate (per unit volume) from the ground state and averaged over $\chi$ phase space as a function of  density in the outer crust  using the single-nucleus table from \cite{chamel}. Curves plotted with solid, dashed and dash-dotted lines correspond to excitation of phonons with $|\vec{k}|\rightarrow 0$ for $m_\chi=500, 100$ and $5$ MeV and $n_\chi/n_{0,\chi}=10$. We also plot for the sake of comparison the specific excitation rate at $|\vec{k}|\rightarrow 0$, $R_{\nu 0}$, for neutrinos with masses $m_{\nu}=0.1, 1$ eV with dotted and doble-dashed lines, respectively. Note, however,  that in this later case, there is a strong momentum dependence that declines rapidly. We can fit this behavior for $m_{\nu}=0.1$ eV as $ R^0_\nu(|\vec{k}|)=R_{\nu 0} e^{ \left(\frac{ -1754|\vec{k}|}{1 \,\rm eV}\right)}$ and for $m_{\nu}=1$ eV as $ R^0_\nu(|\vec{k}|)=R_{\nu 0} e^{-\left( \frac{ 2561.3|\vec{k}|}{1 \,\rm eV}\right)}$.

We have verified that since, typically, the speed of the thermalized LDM particles far from the star is essentially $v_\chi \sim v_\infty\sim 10^{-3}$, when hitting the NS it has already acquired a boosted energy. Using an estimate based on a monochromatic value $E_{\chi}=\gamma_{NS} m_{\chi}$, $\gamma_{NS}=1.26$ we can  straightforwardly integrate and obtain the analytical result
\begin{equation}
R^{(0)}_{{k}}=\frac{n_\chi n^2_A V}{4 (2 \pi)^3 m_{\chi}^3m_Ac_l}\frac{|\gamma_{NS} m_{\chi}-|\vec{k}|c_l|}{\sqrt{\gamma^2_{NS}-1}}a^2,
\label{ro}
\end{equation}
that under-predicts the exact result by $\sim 20\%$.  As deduced from the previous expression  eq.(\ref{ro}) the rate is indeed constant as a function of momentum as the inequality $\gamma_{NS} m_{\chi} \ll |\vec{k}|c_l|$ is fulfilled. It seems that the contribution of the phase space distribution of LDM may also have strong impact on the results, as it happens for the Sun or Earth.  

\begin{figure}[ht]
\centering
\includegraphics[width=0.5\textwidth, angle=0,scale=1.25]{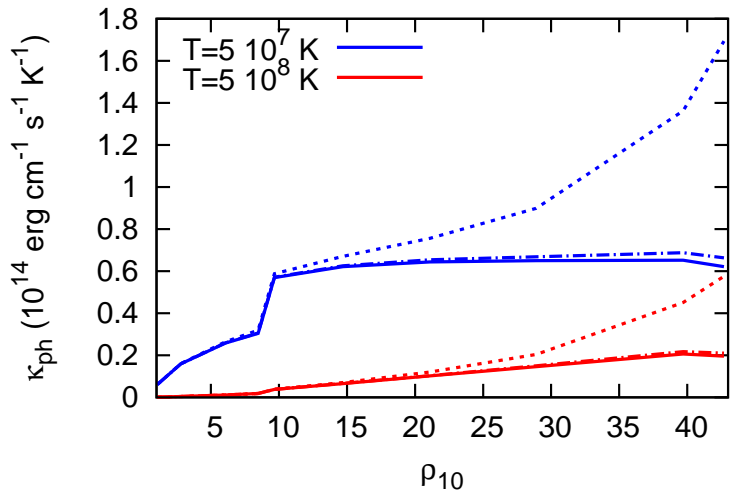}
\caption{Phonon thermal conductivity as a function of density (in units of $10^{10}$ $\rm g/cm^3$) for temperatures $T=5\,10^7$ K (blue), $5\,10^8$ K (red) and $m_\chi=100$ MeV. Dash-dotted and dashed lines depict the impact of a LDM density $n_\chi/n_{0\chi}=10, 100$. Solid lines are the standard thermal result with no DM for each case. See text for details.}
\label{Fig2}
\end{figure}

\section{Astrophysical impact on thermal conductivity}

Phonon production  can be crucial for determining further transport properties, in particular, thermal  conductivity in an ion-electron system such as that in the outer NS crust. As an important contribution to the total ion conductivity, $\kappa_i$, partial ion conductivities due to ion-ion, $\kappa_{ii}\equiv \kappa_{ph}$, and ion-electron collisions, $\kappa_{ie}$, must be added \cite{chugunov} under the prescription  $\kappa^{-1}_i=\kappa^{-1}_{ii}+\kappa^{-1}_{ie}$. Standard mechanisms to produce lattice vibrations include thermal excitations, as analyzed in detail in previous works \cite{pot, baiko}. In a NS, the outer crust  can be modelled under the one-component-plasma description. This low density solid phase can be classified according to the Coulomb coupling parameter $\Gamma=Z^2e^2/a k_B T$ where $a=(4 \pi n_A/3)^{1/3}$ is the ion sphere radius. It is already known that typically for $\Gamma \ge \Gamma_m\simeq 175$, or below melting temperature $T<T_m$, single-ion systems crystallize \cite{gamma}.

There are a number of processes that can affect thermal conductivity in the medium. The so-called U-processes \cite{ziman} are responsible for  modifying the electron conductivity such that for high temperatures, $T>T_U$,  electrons move almost freely. Assuming a bcc lattice, $T_U\simeq 0.07T_D$. Thus in the scenario depicted here,  the  temperature range must be $T_U <T<T_D<T_m$ for each density considered.  According to kinetic theory, the thermal conductivity can be written in the form \cite{ziman} 
\begin{equation}
\kappa_{ii}=\frac{1}{3} k_B C_A n_A c_l L_{ph},
\end{equation} 

where $C_A=9 \left(\frac{T}{T_D}\right)^3 \int_0^{T_D/T} \frac{x^4 e^x dx}{(e^x -1)^2}$ is the phonon (dimensionless) heat capacity {\it per ion}, $L_{ph}$ is an effective  phonon mean free path that includes all scattering processes considered:  U-processes and impurity (I) scattering processes (both dissipative) and the phonon normal (N) scattering which are non dissipative $L^{-1}_{ph}=L^{-1}_U+L^{-1}_I+L^{-1}_N$. Typically the thermal conductivity is related to the thermal phonon number  at temperature $T$, $L_{ph}\sim 1/N_{0,k \lambda}$ where $N_{0,k \lambda}=(e^{\omega_{k\lambda}/k_B T}-1)^{-1}$. The contribution from DM can be obtained by the net number of phonons that results from the competition of thermal and scattering excitation and stimulated emission \cite{ferreras} in a 4-volume $\delta V \delta t$ using the averaged rate per unit volume and  weighting with the incoming distribution providing the frequencies of different values of momenta we obtain

\begin{equation}
N_{k\lambda}\simeq N_{0,k\lambda}+ R^{(0)}_k \delta V \delta t - \int \frac{d^3\vec{p}}{n_\chi} \,f_{\chi}(\vec{p}) {\tilde R^{(0)}_k} N_{0,k\lambda} e^{(\omega_{k,\lambda}+ {\vec k}.{\vec v})/ K_\chi} \delta V \delta t,
\end{equation}
 where $K_\chi=(\gamma-1)m_\chi$ is the $\chi$ kinetic energy and ${\tilde R^{(0)}_k}$ is the single phonon excitation rate for each particular momentum value (not averaged over incoming $\chi$ momenta). Since the source (NS) is in relative motion to the LDM flux, there is a Doppler shift characterized by the source velocity ${v}\equiv v_{NS}\sim 10^{-2}$ i.e galactic NS drift velocity. 
The distribution of NS in our galaxy peak at distances $\langle r \rangle_{max} \lesssim 4$ Kpc \cite{lorimer} where the DM density is enhanced with respect to the solar neighborhood value $n_{0,\chi}\simeq 0.3$ $\rm GeV/cm^3$ thus we will consider $n_\chi\simeq (10, 100)n_{0,\chi}$ as prescribed by popular galactic DM distribution profiles.
\begin{figure}[ht]
\centering
\includegraphics[width=0.5\textwidth, angle=0,scale=1.25]{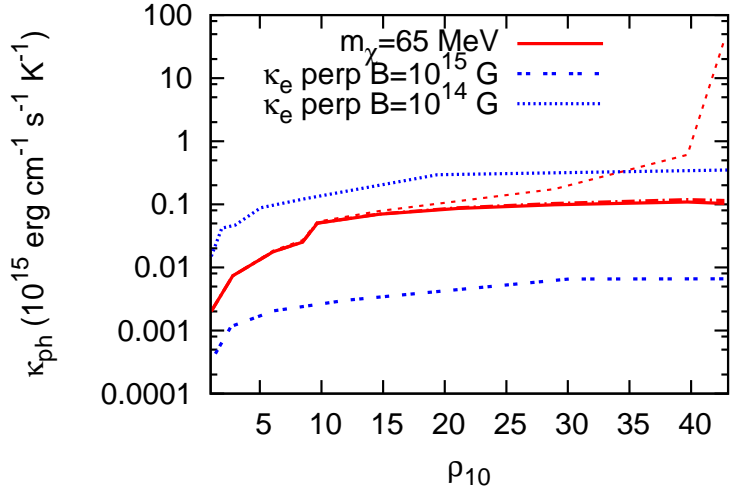}
\caption{Phonon thermal conductivity as a function of density (in units of $10^{10}$ $\rm g/cm^3$) at  $T=10^8$ K and $m_\chi=65$ MeV. Solid, dot-dashed and dashed lines correspond to cases with no DM, $n_\chi/n_{0,\chi}=10,\,100$. Perpendicular electron thermal conductivity is also shown for $B=10^{14}, 10^{15}$ G.}
\label{Fig3}
\end{figure}
In Fig. (\ref{Fig2}) we show the phonon thermal conductivity as a function of density (in units of $10^{10}$ $\rm g/cm^3$) at $T=5\,10^7 \,$ K, $5\,10^8$ K typical for the base of the crust, for $m_\chi=100$ MeV. Solid  lines are the standard thermal result with no DM. Dash-dotted and dashed lines correspond to $n_\chi/n_{0,\chi}=10, \,100$, respectively. 
We  see that at the largest LDM local densities considered, there is a clear enhancement over the thermal result well inside the outer crust. This corresponds to the site where the DM-induced effects have the most influence \cite{azorin} as this is the most massive part of the outer crust. Below these densities, there is a negligible change though. At lower $T$, the effect of a perturbation over the thermal phonon population is more important. Enhanced (decreased) conductivities at moderate LDM densities are due to a net reduction (increase) of the number of phonons in the lattice as a result of cancellation of modes. As a representative  scenario, we have taken $|\vec{k}|=0.01/a$ at each density since we have verified that this choice verifies the kinematical restrictions on $|\vec{k}|$ when performing the averages over phase space distribution as discussed in section II. Besides, rates are mostly constant at low $|\vec{k}|$. Note that in standard calculations \cite{chugunov}, there is no momentum dependence as they replace the  frequency mode $\omega_{k\lambda}$ by a constant threshold. We must bear in mind that this result must be compiled with a realistic impurity fraction so that conductivity remains finite. We have considered $L_I\sim 5a$ \cite{chugunov}.

In order to understand the significance  of our result in the dense stellar context in Fig.(\ref{Fig3}) we show the phonon thermal conductivity as a function of density (in units of $10^{10}$ $\rm g/cm^3$) at  $T=10^8$ K and $m_\chi=65$ MeV for $|\vec{k}|=0.01/a$. Solid, dot-dashed and dashed lines correspond to cases with no DM, $n_\chi/n_{0,\chi}=10,\,100$, respectively. Electron thermal conductivity is also shown for magnetized realistic scenarios  in the perpendicular direction to a magnetic field $B$ of strength $B=10^{14}$ G (dotted) and $B=10^{15}$ G (doble dotted). Ions are mostly not affected by the presence of a magnetic field. The parallel direction electronic contribution is not depicted here since it is typically much larger $k_{e \parallel}\sim 10^{17}-10^{19}$ $\rm erg\,cm^{-1} \,s^{-1}\, K^{-1}$.  Since we perform averages over the $\chi$ phase space we again use $|\vec{k}|=0.01/a$. On the plot we can see that the electronic contribution in the perpendicular direction falls below the enhanced $n_\chi/n_{0,\chi}=100$ DM value for densities $\gtrsim 3.5\,10^{11}$ $\rm g/cm^3$. Note that the low value chosen for $|\vec{k}|$ in this plot is to be understood as a compromise value, larger $|\vec{k}|$ values would imply the impossibility of exciting phonons from low-momenta incoming LDM. Since the global conductivity is $\kappa=\kappa_e+\kappa_{ph}$, the obtained result is expected to contribute to the reduction of the difference in heat conduction in both directions and thus to the isotropization of the NS surface temperature pattern as seen in \cite{azorin}  for standard physics. Temperatures would be  smoothly driven towards more isothermal profiles for latitudes among pole and equator. It is already known  \cite{kamin} that the outer crust plays an important  role in regulating the relation among temperature in the base of it and the surface. The detailed calculation of this implication for surface temperatures remains, however, for future work.

\section{Conclusions}

In conclusion, we have derived for the first time the single phonon excitation rate in the outer NS crust for  relativistic LDM particles in the sub-GeV mass range. We have found that this rate is constant with the phonon momentum and much larger than for cosmological neutrinos at finite $|\vec{k}|$. A non-negligible correction to the local phonon excitation rate of $\sim 20\%$ is obtained when full relativistic phase distribution functions are considered for the incoming $\chi$ particles with respect to a monochromatic approximation, that under-predicts the result.

As an astrophysical consequence of the previous, we have calculated the ion thermal conductivity in the dense and hot outer envelope founding that it can be largely enhanced at LDM densities in the maximum of the NS galactic distribution  $n_\chi\sim 100 n_{0,\chi}$ due to a net modification of the acoustic phonon population. This effect is non negligible at  densities beyond $\sim 3.5\, 10^{11}$ $\rm g/cm^3$ in the base of the outer crust at the level of standard ion-electron or thermal effects \cite{chugunov, pot}.  We do not expect the degenerate electron contribution to largely modify this result as this would  mildly screen nuclear charge in the lattice however it remains to be further studied. 
Although a detailed study of the quantitative effect in the surface temperature pattern remains to be undertaken, it is expected that for magnetized NSs the  LDM-enhanced global  enhancement of the perpendicular thermal conductivity allows a reduction of the difference of heat transport among parallel and perpendicular directions to  the magnetid field. Based on previous works only including standard thermal contributions we expect that, as a natural consequence, the surface temperature profile would be more isotropic yielding flatter profiles for intermediate latitudes and remains to be calculated in a future contribution.

\section{Acknowledgments}

We acknowledge  useful comments from J. Pons and C. Albertus. This research has been partially supported by MULTIDARK and FIS2015-65140 MINECO projects and at IAP by the ERC project  267117 (DARK) hosted by Universit\'e Pierre et Marie Curie - Paris 6   and at JHU by NSF grant OIA-1124403. M. Cerme\~no is supported by a fellowship from the University of Salamanca.



\begin{thebibliography}{9}
%
\bibitem{cdm} Planck Collaboration, P. Ade et al., Astron. Astrophys. 571 (2014) A16.
\bibitem{dama} DAMA Collaboration, LIBRA Collaboration, R. Bernabei et al., Eur. Phys. J. C 67 (2010) 39, arXiv:1002.1028.
\bibitem{lin} T. Lin, H. Yu and K. M. Zurek,  Phys. Rev. D 85 (2012) 063503, arXiv: 1111.0293v2 [hep-ph].

\bibitem{ele} R. Essig, M. Fern\'andez Serra, J. Mardon et al, arXiv: 1509.01598v2[hep-ph].
\bibitem{chamel} S. B. Ruster, M. Hempel, and J. Schaffner-Bielich, Phys. Rev. C, 73,  (2006) 035804.
\bibitem{pasta} C.J. Horowitz, M.A. P\'erez-Garc\'ia, J. Piekarewicz, Phys. Rev. C. 69 (2004) 045804; M. A. P\'erez-Garc\'ia, J. Math. Chem. 48 (2010) 21.

\bibitem{ziman} Ziman J. M., Electrons and Phonons, Oxford Univ. Press, Oxford (1960).

\bibitem{carr} W. J. Carr, Phys. Rev., 122 (1961) 1437.

\bibitem{ferreras}  I. Ferreras and I. Wasserman, Phys. Rev. D 52 (1995) 5459; R. Chandrasekharan, Ph. D. Thesis (2003).

\bibitem{haxton}A. L. Fitzpatrick, W. Haxton, E. Katz, N. Lubbers and Y. Xu, arXiv:1203.3542v3 [hep-ph].

\bibitem{cermeno} M. Cerme\~no, M. A. P\'erez-Garc\'ia, J. Silk, Phys. Rev. D 94 (2016) 023509, arXiv:1511.04071v1 [hep-ph].

\bibitem{tait} J. Goodman, M. Ibe, A. Rajaraman, W. Spherd, T.M. P. Tait, and H. Yu, Phys. Rev. D. 82 (2010) 116010.

\bibitem{ci} K. Cheung, P. Tseng, Y. S. Tsai, T. Yuan, JCAP 05 (2012) 001.

\bibitem{limits} CMS Collaboration (V. Khachatryan, et al.) Eur. Phys. J. C 75 (2015) 235 [arXiv:1408.3583]; G. Aad, et al., Phys. Rev. D 90 (2014) 012004; G. Aad, et al., Phys. Rev. D 91 (2015) 012008.

\bibitem{ddetec} J. Zheng et al.,  Nucl. Phys. B 854 (2012) 350, arXiv 1012.2022v3. 

\bibitem{gluscev}V. Gluscevic, M. I. Gresham, S. D. McDermott, A. H. G. Peter and K. M. Zurek, JCAP 12 (2015) 057, arXiv:1506.04454v2 [hep-ph].

\bibitem{Ashcroft} N. W. Ashcroft and N. D. Mermin,  Solid State Physics, Saunders College, Philadelphia (1976).

\bibitem{horo} C. J. Horowitz, arXiv:1205.3541v1 [astro-ph.HE].

\bibitem{dir} A. M. Green, Phys. Rev. D 66 (2002) 083003; ibid, Mod. Phys. Lett. A, 27  (2012) 1230004.

\bibitem{ind} M. Fornasa and A. M. Green, Phys. Rev. D 89 (2014)  063531.

\bibitem{choi} K. Choi, C. Rott and Y. Itow, JCAP 05 (2014) 049.

\bibitem{hedman} M. M. Hedman, JCAP 09 (2013) 029, arXiv:1307.0652v2 [astro-ph.CO].

\bibitem{chugunov} A. I. Chugunov and P. Haensel, Mon. Not. R. Astron. Soc. 381,  (2007) 1143.

\bibitem{pot} J. W. Negele, and D. Vautherin, Nucl. Phys. A, 207 (1973) 298; A. Y. Potekhin, G. Chabrier and D. G. Yakovlev, 323 A\&A (1197), 415.
\bibitem{baiko} D. A. Baiko, A. D. Kaminker, A. Y. Potekhin, and D. G. Yakovlev, Phys. Rev. Lett. 81 (1998) 5556.
\bibitem {gamma} H. Nagara, Y. Nagata, and T. Nakamura, Phys. Rev. A 36 (1987) 1859.

\bibitem{juttner} Juttner F., Das Maxwellsche Gesetz der Geschwindigkeitsverteilung in der Relativtheorie, Ann. Physik und Chemie, 34, (1911) 856.

\bibitem{Kremer} G. M. Kremer, J. Stat. Mech. (2013) P04016, arXiv:1212.5573v4.

\bibitem{Hakim} R. Hakim, Introduction to Relativistic Statistical Mechanics Classical and Quantum, World Scientific (2011).
\bibitem{Cercignani} C. Cercignani et al., The Relativistic Boltzmann Equation: Theory and Applications, Birkh\"auser Verlag (2002).

\bibitem{lorimer} D. R. Lorimer, A. J. Faulkner,A. G. Lyne et al., Mon. Not. R. Astron. Soc. 372, (2006) 777.
\bibitem{azorin} J. F. P\'erez-Azor\'in, J. A. Miralles, and J. A. Pons, A\&A 451, (2006) 1009.
\bibitem{kamin} A. D. Kaminker  et al, Astrophys. Space Sci. 308 (2007) 423.


\end{thebibliography}
\end{document}